# Self-Similar Evolution of Cosmological Density Fluctuations

Bhuvnesh Jain[1,2] and Edmund Bertschinger[1]

[1]Department of Physics, MIT, Cambridge, MA 02139 USA

[2]Max-Planck-Institut für Astrophysik, Karl-Schwarzschild-Strasse 1, 85748 Garching, Germany; bjain@mpa-garching.mpg.de

## ABSTRACT

The gravitational evolution of scale free initial spectra $P(k) \propto k^n$ in an Einstein-de Sitter universe is widely believed to be self-similar for $-3 < n < 4$. However, for $-3 < n < -1$ the existence of self-similar scaling has not been adequately demonstrated. Here we investigate the possible breaking of self-similar scaling due to the nonlinear contributions of long wave modes. For $n < -1$ the nonlinear terms in the Fourier space fluid equations contain terms that diverge due to contributions from wavenumber $k \to 0$ (the long wave limit). To assess the possible dynamical effects of this divergence the limit of long wave contributions is investigated in detail using two different analytical approaches.

Perturbative contributions to the power spectrum are examined. It is shown that for $n < -1$ there are divergent contributions at all orders. However, at every order the leading order divergent terms cancel out exactly. This does not rule out the existence of a weaker but nevertheless divergent net contribution. The second approach consists of a non-perturbative approximation, developed to study the nonlinear effects of long wave mode coupling. A solution for the phase shift of the Fourier space density is obtained which is divergent for $n < -1$. A kinematical interpretation of the divergence of the phase shift, related to the translational motion induced by the large-scale bulk velocity, is given. Our analysis indicates that the amplitude of the density is *not* affected by the divergent terms and should therefore display the standard self-similar scaling. Thus both analytical approaches lead to the conclusion that the self-similar scaling of physically relevant measures of the growth of density perturbations is preserved.

*Subject headings:* cosmology: theory — large-scale structure of universe — galaxies: clustering — galaxies: formation

## 1. Introduction

The self-similar scaling of density perturbations with scale free initial conditions in a spatially flat universe has been a useful theoretical tool in studying structure formation. It has been widely used to study gravitational clustering in cosmology and has been tested by several studies using N-body simulations. However self-similar scaling for scale free initial spectra $P(k) \propto k^n$ has not been adequately demonstrated for $n < -1$ because the requirements of dynamic range get increasingly difficult to meet as $n$ gets smaller. Indeed results of some two dimensional studies suggest a breaking of self-similar scaling for $n = -2$ in three dimensions. Analytical analyses have been limited to the observation that the linear peculiar velocity field diverges for $n < -1$, but the linear density contrast does not diverge provided $n > -3$. This would suggest that while there may be formal problems with establishing self-similarity for $n < -1$, in practice it should hold as long as $n > -3$.

Our goal in this paper is to analyze the dynamics of the coupling of long wave modes by analytical techniques to address the question of whether self-similar scaling is broken for scale free spectra with $n < -1$. A point that will be highlighted in the following sections is that the answer can depend on the particular statistic used to pose the question. The question of real interest concerns the self-similar growth of measures of density perturbations that relate to the formation of structure. Therefore our attempt will be to identify such statistical measures and examine their scaling behavior. An analysis of scale free N-body simulations that addresses the same issues is presented in a following paper (Bertschinger & Jain 1995, Paper II).

Section 1.1 gives a detailed discussion of the concept of self-similar scaling and its application to cosmology. The statistical nature of self-similar scaling is emphasized and an account of previous work that motivated our study is given. This is also done with a view to anticipating some of the subtler issues relating to a possible breakdown of self-similarity that emerge in our subsequent analysis. Section 2 provides an assessment of whether $-3 < n < -1$ is expected to yield self-similar evolution on the basis of simple dynamics. We argue that the issue can only be settled by a full consideration of the dynamical coupling of long wave modes, rather than by studying the convergence of particular statistics using linear solutions. In our analysis we work with the Fourier space density field as it quantifies the relative amounts of power on different scales most directly. In Section 3 we use perturbation theory to study self-similar scaling. We begin in Section 3.1 by formulating perturbation theory in a way that obeys this scaling at every order provided there are no long wave divergences. In Section 3.2 we demonstrate that there are potentially divergent perturbative contributions to the power spectrum, but the leading order contributions exactly cancel out. Section 4 presents an alternative, non-perturbative approximation to estimate the coupling of long wave modes. The evolution of the amplitude and phase of Fourier modes is studied separately and a solution for the rms phase shift is presented. Kinematical effects which do not affect perturbation growth are distinguished from dynamical ones which do in order to assess the scaling of the amplitude of the density. We conclude in Section 5.

## 1.1. Self-Similarity and Structure Formation in Cosmology

A physical system is expected to display self-similar evolution if there is no preferred scale in the system, either in the initial conditions or in its dynamical behavior. The differential equations governing the evolution of such a system generally admit of a self-similar solution. Suppose the basic evolution equation is a partial differential equation for the phase space density $f(\vec{x}, \vec{p}, t)$, where $\vec{x}$ is the spatial position, $\vec{p}$ is the momentum, and $t$ is the time variable. In a self-similar system it is possible to re-cast the equation in a form with a solution $f = t^\alpha \hat{f}(\vec{x}/t^\beta, \vec{p}/t^\gamma)$, where $\hat{f}$ is in general an unknown function. If the constant coefficients $\alpha$, $\beta$, and $\gamma$ are known then the time dependence of $f$ is present only through the rescaled $\vec{x}$ and $\vec{p}$ coordinates, aside from the overall factor of $t^\alpha$. This special form of the solution is defined to be self-similar: the phase space density at time $t_2$ is related to that at time $t_1$ as

$$f(\vec{x}_2, \vec{p}_2, t_2) = \left(\frac{t_2}{t_1}\right)^\alpha f(\vec{x}_1, \vec{p}_1, t_1), \qquad (1)$$

where $\vec{x}_1 = \vec{x}_2 (t_1/t_2)^\beta$, and $\vec{p}_1 = \vec{p}_2 (t_1/t_2)^\gamma$. Equation (1) explicitly demonstrates that the phase space density for any $(\vec{x}, \vec{p})$ at all times $t_2$ can be obtained merely by re-scaling from some chosen time $t_1$. Clearly self-similarity is a powerful constraint because any statistical measure constructed from the the phase space density should be described by the appropriate scaling of coordinates consistent with equation (1).

We now consider the similarity properties of gravitational dynamics in a zero-pressure Einstein-de Sitter cosmology. An Einstein-de Sitter universe refers to the model with the cosmological density parameter $\Omega = \Omega_{matter} = 1$ and zero cosmological constant, so that the universe is spatially flat. The gravitational interaction also does not pick a special length scale. Further let the initial power spectrum be a power law, $P(k) \propto k^n$, over length scales of interest. Thus so far there is no preferred length scale in the system. The amplitude of the power spectrum can be used to define a characteristic physical length scale: the scale at which the rms smoothed density contrast equals unity is the conventional choice. To within an order of magnitude it is the scale at which over-densities collapse out of the background expansion. The presence of this scale does not intereferewith self-similar scaling, rather it provides the reference scale required for scaling the spatial variable according to the similarity solution.

The explicit similarity transformation for the single particle phase space density $f(\vec{x}, \vec{p}, t)$ is described in Section 73 of Peebles (1980). It is also shown that knowing the linear solution is sufficient to fix the indices $\alpha$, $\beta$ and $\gamma$ in terms of the spectral index $n$ of the initial spectrum. The resulting self-similar scaling of spatial length scales $x$, and wavenumber scales $k$ is:

$$x_{ss}(t) \propto a(t)^{2/(3+n)} \quad ; \quad k_{ss}(t) \sim x_{ss}(t)^{-1} \propto a(t)^{-2/(3+n)}. \tag{2}$$

The similarity solution for $f$ is obtained by dimensional analysis of the differential equation describing its evolution. Whether or not the solution applies depends on the details of the initial conditions. A popular choice for the initial fluctuations in cosmology is that of a Gaussian random field which is statistically homogeneous and isotropic in space. For a given realization, the stochasticity of the initial distribution in space precludes the similarity solution for $f$ from being valid. Although the statistical properties of the distribution are scale free for a power law initial spectrum, the distribution in any one realization is not spatially self-similar, or independent of the spatial scale in any meaningful way. This means that the phase space density $f$ for a particular realization cannot obey the similarity solution. All the results on self-similar scaling that will be discussed below are therefore valid only for ensemble averages (averages across different independent realizations). This distinguishes the self-similar scaling of cosmological perturbations from standard examples of self-similar systems.

The ensemble averages of $f$ or products of $f$ do evolve self-similarly because ensemble averaging removes the stochastic character of the initial conditions. Thus the self-similar scaling for the 2−point correlation function $\xi(x,t)$ can be obtained from the formal solution for $f$, and it is valid even though the self-similar solution for $f$ does not hold. Its validity can be verified by dimensional analysis of the evolution equation for $\xi$ that is obtained by taking moments of the BBGKY hierarchy equations (e.g., equation 1.1 of Peebles 1980). The solution for the power spectrum is obtained by Fourier transforming $\xi(x,t)$, and is

$$P(k,t) = a^{3\alpha} k_0^{-3} \hat{P}(ka^\alpha/k_0) \;, \tag{3}$$

where $\alpha = 2/(3+n)$, $k_0$ is a constant which must be determined from the initial conditions, and $\hat{P}$ is an unspecified dimensionless function. It is easy to verify that the linear spectrum $P_{11}(k,\tau) \propto a^2 k^n$ is consistent with this functional form.

Likewise the scaling of all statistical measures defined as ensemble averages of products of $f$ (and of their momentum moments) can be straightforwardly determined. Using the ergodic theorem the solutions for ensemble averages can then be applied to averages over sufficiently large volumes in space. A spatial statistic which follows the self-similar solution is a function of the spatial variable scaled by a power of time, rather than of time and space separately. This provides for a self-similarity in time (in this statistical sense) in the evolution of structure.

The discrete nature of particles (N-body or galaxies) introduces a scale, namely the mean interparticle separation, which breaks the idealized self-similar scaling of a perfect fluid. Such a departure from perfect self-similarity is typical of all realistic physical systems. The notion of intermediate asymptotic self-similarity, i.e., self-similar scaling over a restricted range of parameters, is used in such situations (Barenblatt 1979). In the cosmological context it simply means that the range of length scales over which self-similar scaling is accurately followed are restricted to be sufficiently larger than the interparticle separation (and in the case of N-body simulations, the force resolution scale).

Intermediate asymptotic self-similarity is a useful property even for realistic cosmological spectra like the CDM spectrum which are not scale free. The physical processes at work in the radiation dominated era imprint characteristic length scales on the spectrum. These spectra are nevertheless approximate power laws on a restricted range of $k$, over which their evolution may be well described by the similarity solution for the corresponding scale free spectrum. Thus the CDM spectrum has $n \simeq -2$ on galactic scales and $n \simeq -1$ on cluster/supercluster scales; therefore the study of scale free spectra with $-1 \lesssim n \lesssim -2$ is relevant for understanding the development of large scale structure in a CDM-like model. Moreover, recently the idea of self-similar scaling has been extended to provide a more direct scaling description of the evolution of CDM-like initial spectra (Hamilton et al. 1991; Peacock & Dodds 1994; Mo, Jain & White 1995).

An aspect of self-similarity which merits attention is the range of $n$, the spectral index of the initial spectrum, for which the statistics characterizing the growth of perturbations are well defined. More precisely, $n$ must be restricted from below and above to prevent statistical measures of interest from diverging as the size of the system is made infinitely large and the interparticle spacing made infinitesimally small, respectively. In a finite system such a divergence is manifested by the influence of the largest (or smallest) scales on the evolution of all intermediate scales of the system. This occurs if either the statistic used is ill defined even in the initial configuration, or the dynamical influence of increasingly large or small scales is unbounded. If the former is true in an otherwise reasonable initial configuration, then it must mean that the particular statistic is not a suitable measure of the properties of the system (similar to the case of well behaved probability distributions having divergent moments). However, if the breaking of self-similarity is due to a divergent dynamical effect in a statistic of interest then it bears closer examination. The goal of this work is to examine the possible breaking of self-similar evolution for power law initial spectra with a view to assessing its influence on the formation of structure.

Early studies of self-similar evolution in cosmology include those of Peebles (1974); Press & Schechter (1974); Davis & Peebles (1977); and Efstathiou & Eastwood (1981). Davis & Peebles (1977) made a detailed analysis of the BBGKY hierarchy equations and presented approximate solutions for the deeply nonlinear regime based on the stable clustering ansatz. Efstathiou et al. (1988) tested self-similar scaling in N-body simulations of scale free spectra with $n = -2, -1, 0, 1$. They examined the scaling of the correlation function $\xi(x, t)$, and of the multiplicity function describing the distribution of bound objects. They verified the predicted scaling for both statistics, and found consistency with the picture of hierarchical formation of nonlinear structure on increasingly large length scales. Their results for $n = -2$ did not match with the self-similar scaling as well as the other cases. Bertschinger & Gelb (1991) used better resolution simulations to address these questions and also found similar results. These authors concluded that the reason for the weakness of the $n = -2$ results was the finite size of their simulation box, as the $n = -2$ case has more power on large scales and therefore requires a larger box-size to approximate the infinite volume limit with the same accuracy as larger values of $n$.

Recently Lacey & Cole (1994) have examined the self-similar scaling of the number density of nonlinear clumps for scale-free spectra. Their results indicate that self-similar scaling works reasonably well for the statistics they measure, even for $n = -2$. Mo, Jain & White (1995) reach the same conclusion for the correlation function and power spectrum. However, in all the above studies the self-similar scaling of an averaged quantity is tested without making a comparison with alternate scalings. This makes it difficult to distinguish the effect of limited numerical resolution from a real breakdown of self-similar scaling. In Paper II we demonstrate that even with the current state-of-the-art simulations it is extremely difficult to use an averaged quantity like the power spectrum to critically test self-similar scaling for $n = -2$.

The N-body results of Ryden & Gramann (1991), and Gramann (1992) suggest that the $n = -2$ case is different for a fundamental reason. They studied $n = -1$ simulations in two dimensions, which are the analog of $n = -2$ in three dimensions, and examined the scaling of the phase (Ryden & Gramann 1991), and then both phase and amplitude (Gramann 1992) of the Fourier transform of the density field. The scaling was found to be different from the standard self-similar scaling. Characteristic wavenumber scales, instead of following the self-similar scaling, given in two dimensions by $k_{ss}(t) \propto a(t)^{-2/(2+n)} \propto a(t)^{-2}$, showed the scaling $k \propto a(t)^{-1}$. Other studies in two dimensions also suggest that a transition in nonlinear evolution occurs at $n = -1$ (Klypin & Melott 1992, and references therein). Motivated by Gramann's results we had re-examined the $n = -2$ simulation presented in Bertschinger & Gelb (1991) and found that the results were ambiguous, and that a bigger simulation would be needed to provide a definitive answer. This had provided the initial motivation for our analytical investigation as well.

It has been noted all along in the literature that the formal bounds on $n$ for the self-similar solution to be applicable are $-1 < n < 1$. The requirements of an upper (lower) limit are made to prevent the single particle velocity dispersion from diverging due to contributions from small (large) length scales. These bounds on $n$ are clearly stated as the domain of applicability of self-similar scaling in Peebles & Davis (1977), Efstathiou et al. (1988), and in the recent review of Efstathiou (1990). However, it appears to be implicitly believed that self-similar scaling is applicable for $n > -3$, rather than $n > -1$. This is probably because the divergence of the bulk velocity field need not affect the growth of perturbations. The primary quantity that measures perturbation growth is the rms density contrast which is indeed convergent for $n > -3$ as $k \to 0$. Thus Peebles (1993, p. 545) presents the standard self-similar scaling as being applicable for $-3 < n < 4$ (increasing the upper limit to $n = 4$ relies on the asymptotic behavior of second order contributions to the density). Efstathiou (1990) was more cautious, but nevertheless hoped that: "If $n$ lies outside this range (i.e., $-1 < n < 1$), the clustering may still approximate self-similar evolution over restricted ranges of length and time, although $n > -3$ is required to ensure that clustering proceeds from small to large scales."

## 2. Long Wave Divergences for $n < -1$

We suppose for simplicity that the matter distribution after recombination may be approximated as a pressureless fluid with no vorticity. We further assume that peculiar velocities are nonrelativistic and that the wavelengths of interest are much smaller than the Hubble distance $cH^{-1}$ so that a nonrelativistic Newtonian treatment is valid. Using comoving coordinates $\vec{x}$ and conformal time $d\tau = dt/a(t)$, where $a(t)$ is the expansion scale factor, the nonrelativistic cosmological fluid equations are

$$\frac{\partial \delta}{\partial \tau} + \vec{\nabla} \cdot [(1+\delta)\vec{v}] = 0 , \tag{4a}$$

$$\frac{\partial \vec{v}}{\partial \tau} + \left(\vec{v} \cdot \vec{\nabla}\right) \vec{v} = -\frac{\dot{a}}{a} \vec{v} - \vec{\nabla} \phi , \tag{4b}$$

$$\nabla^2 \phi = 4\pi G a^2 \bar{\rho} \delta , \tag{4c}$$

where $\dot{a} \equiv da/d\tau$. Note that $\vec{v} \equiv d\vec{x}/d\tau$ is the proper peculiar velocity, which we take to be a potential field so that $\vec{v}$ is fully specified by its divergence:

$$\theta \equiv \vec{\nabla} \cdot \vec{v} \ . \tag{5}$$

We assume an Einstein-de Sitter ($\Omega = 1$) universe, with $a \propto t^{2/3} \propto \tau^2$. We will also assume that the initial (linear) density fluctuation field is a gaussian random field.

To quantify the amplitude of fluctuations of various scales it is preferable to work with the Fourier transform of the density fluctuation field, which we define as

$$\hat{\delta}(\vec{k},\tau) = \int \frac{d^3x}{(2\pi)^3} \, e^{-i\vec{k}\cdot\vec{x}} \delta(\vec{x},\tau) \ , \tag{6}$$

and similarly for $\hat{\theta}(\vec{k},\tau)$. The power spectrum (power spectral density) of $\delta(\vec{x},\tau)$ is defined by the ensemble average two-point function,

$$\langle \hat{\delta}(\vec{k}_1,\tau) \hat{\delta}(\vec{k}_2,\tau) \rangle = P(k_1,\tau) \, \delta_{\rm D}(\vec{k}_1 + \vec{k}_2) \ , \tag{7}$$

where $\delta_{\rm D}$ is the Dirac delta function, required for a spatially homogeneous random density field. For a homogeneous and isotropic random field the power spectrum depends only on the magnitude of the wavevector. The contribution to the variance of $\delta(\vec{x},\tau)$ from waves in the wavevector volume element $d^3k$ is $P(k,\tau) d^3k$.

Fourier transforming equations (4) gives:

$$\frac{\partial \delta}{\partial \tau} + \theta = -\int d^3k_1 \, \frac{\vec{k}\cdot\vec{k}_1}{k_1^2} \, \theta(\vec{k}_1,\tau)\,\delta(\vec{k}-\vec{k}_1,\tau) \ , \tag{8a}$$

$$\frac{\partial \theta}{\partial \tau} + \frac{\dot{a}}{a}\theta + \frac{6}{\tau^2}\delta = -\int d^3k_1 \, k^2 \frac{\vec{k}_1\cdot(\vec{k}-\vec{k}_1)}{2k_1^2|\vec{k}-\vec{k}_1|^2} \, \theta(\vec{k}_1,\tau)\,\theta(\vec{k}-\vec{k}_1,\tau) \ . \tag{8b}$$

The fields on the left-hand side are all functions of $\vec{k}$ and $\tau$. The nonlinear terms on the right-hand side of the above equations represent the coupling of modes at all pairs of wavevectors $(\vec{k}_1, \vec{k} - \vec{k}_1)$, which influence the evolution of $\delta$ and $\theta$ at the fixed external wavevector $\vec{k}$.

In order to study the limiting behavior as one of the pair of wavevectors $(\vec{k}_1, \vec{k} - \vec{k}_1)$ approaches 0 in magnitude (i.e., as the wavelength $\lambda = 2\pi/k$ is made infinitely large), consider the variance of the nonlinear terms. For simplicity we take $k_1 \to 0$ in the integral on the right-hand side of equation (8a). Approximating $(\vec{k} - \vec{k}_1)$ by $\vec{k}$, dropping the

dependence on $\tau$, and denoting the resulting variance by $\psi(k)$ we obtain:

$$\psi(k) \simeq \left\langle \int d^3k_1 \frac{\vec{k} \cdot \vec{k}_1}{k_1^2} \theta(\vec{k}_1) \delta(\vec{k}) \int d^3k_2 \frac{\vec{k} \cdot \vec{k}_2}{k_2^2} \theta^*(\vec{k}_2) \delta^*(\vec{k}) \right\rangle,$$

$$= \int d^3k_1 \int d^3k_2 \frac{\vec{k} \cdot \vec{k}_1}{k_1^2} \frac{\vec{k} \cdot \vec{k}_2}{k_2^2} \left\langle \theta(\vec{k}_1) \theta^*(\vec{k}_2) \delta(\vec{k}) \delta^*(\vec{k}) \right\rangle. \quad (9)$$

Now we make the further approximation of taking $k_1$ small enough that the linear solutions are valid, thus giving:

$$\delta(\vec{k}_1, \tau) \simeq a(\tau) \delta_1(\vec{k}_1) \; ; \; \theta(\vec{k}_1, \tau) \simeq -\dot{a}(\tau) \delta_1(\vec{k}_1). \quad (10)$$

Substituting the expression for $\theta(\vec{k}_1, \tau)$ in equation (10) into equation (9) and evaluating the ensemble average using the properties of Gaussian random fields we finally obtain:

$$\psi(k) \simeq \dot{a}^2 \delta_D(0) P(k) \int d^3k_1 \left( \frac{\vec{k} \cdot \vec{k}_1}{k_1^2} \right)^2 P_{11}(k_1) \simeq \dot{a}^2 \delta_D(0) P(k) \, k^2 \int dk_1 P_{11}(k_1). \quad (11)$$

Note that in expressing the 4−point moment of equation (9) in terms of $P(k)$ and $P_{11}(k_1)$ we have assumed that $\delta(\vec{k})$ is a Gaussian random field as well. It is straightfoward to demonstrate that the right-hand side of equation (8b) takes the limiting form shown in equation (11) as well.

Equation (11) indicates that if $P_{11}(k_1) \propto k_1^n$ with $n < -1$, then the right-hand side diverges due to contributions from low $k_1$. Thus a simple examination of the nonlinear terms in the cosmological fluid equations by substituting the initial distribution of the density and velocity field demonstrates the possibility of long wave divergences. These divergences can potentially be present in solutions for $\delta(\vec{k}, \tau)$ and $\theta(\vec{k}, \tau)$ obtained from these equations. It is not possible to make any definitive statements, however, because these are two coupled differential equations — it is necessary to first separate the equations for $\delta$ and $\theta$, and then identify the nonlinear terms that affect the amplitude and phase of each quantity (since they are complex variables) to determine whether the divergent terms affect a particular statistic of interest. This is done in two different ways in Sections 3 and 4.

Before proceeding with a formal analysis of the divergent nonlinear pieces, we make the connection between the divergent nonlinear terms to the advective ($\vec{v} \cdot \vec{\nabla}$) terms in the real space equations. By tracing back the nonlinear terms on the right-hand side of equations (8) to the fluid equations in real space it can be seen that the terms which contribute to equation (9) arise from the $\vec{v} \cdot \vec{\nabla} \delta$ and $\vec{v} \cdot \vec{\nabla} \vec{v}$ terms. It is easy to see why such terms should diverge by examining the relation of the power spectrum of the peculiar velocity to that of the density in linear theory. Using the linear solutions of equation (10) and the definition $\theta(\vec{x}, \tau) = \vec{\nabla} \cdot \vec{v}(\vec{x}, \tau)$, gives,

$$P_{11\,v}(k, \tau) = \dot{a}^2 P_{11}(k)/k^2, \tag{12}$$

where $P_{11\,v}(k)$ is the linear power spectrum of the peculiar velocity. The rms bulk velocity on a scale $x \sim k^{-1}$, $v_b(x, \tau)$, is given by integrating $P_v(k)$ over $k$ with a window function $W(kx)$ (just as one integrates over $P(k)$ for the rms smoothed density contrast):

$$v_b(x, \tau)^2 = \int d^3k P_{11\,v}(k, \tau) W^2(kx) = \dot{a}^2 \, 4\pi \int dk P_{11}(k) W^2(kx). \tag{13}$$

Since $W(kx) \to 1$ as $k \to 0$ (see for example the standard top-hat window function), the integral on the right-hand side of equation (13) diverges at low $k$ for $n < -1$ in the same manner as the integral in equation (11). Thus via the advective ($\vec{v} \cdot \vec{\nabla}$) terms in the fluid equations, the divergence of the nonlinear terms demonstrated in equation (11) can be traced to the bulk velocity field on a given scale receiving divergent contributions from $k \to 0$, i.e., from the long wavelength modes.

We can now understand why this divergence may not affect self-similar scaling: the bulk velocity field does not in general have any influence on the growth of perturbations on small scales. In particular, large contributions to the bulk velocity field from long wave modes correspond to an almost uniform translation of the fluid, and therefore should not couple to the evolution of $\delta$ at all. This reasoning, and indeed the entire analysis of this section, relies on making plausible connections of linearized statistics for $\delta$ and $\vec{v}$ to their nonlinear dynamics. Therefore, while it provides a useful guide to one's intuition, it does not substitute for a rigorous examination of the nonlinear dynamics.

## 3. Self-Similarity and Perturbation Theory

### 3.1. Formalism

The basic formalism of perturbation theory used here most closely follows that of Goroff et al. (1986); it is described in detail in Jain and Bertschinger (1994). We begin by writing the solution to equations (8) as a perturbation series,

$$\hat{\delta}(\vec{k}, \tau) = \sum_{l=1}^{\infty} a^l(\tau) \, \delta_l(\vec{k}) \,, \quad \hat{\theta}(\vec{k}, \tau) = \sum_{l=1}^{\infty} \dot{a}(\tau) a^{l-1}(\tau) \, \theta_l(\vec{k}) \tag{14}$$

It is easy to verify that for $l = 1$ the time dependent part of the solution correctly gives the linear growing modes $\hat{\delta}_1 \propto a(\tau)$ and $\hat{\theta}_1 \propto \dot{a}$ and that the time-dependence is consistent with equations (8) for all $l$. To obtain formal solutions for the $\vec{k}$ dependence at all orders we proceed as follows.

Substituting equation (14) into equations (8) yields, for $l > 1$,

$$l\delta_l(\vec{k}) + \theta_l(\vec{k}) = A_l(\vec{k}), \quad 3\delta_l(\vec{k}) + (1+2l)\theta_l(\vec{k}) = B_l(\vec{k}), \tag{15}$$

where

$$A_l(\vec{k}) \equiv -\int d^3k_1 \int d^3k_2 \, \delta_D(\vec{k}_1 + \vec{k}_2 - \vec{k}) \frac{\vec{k} \cdot \vec{k}_1}{k_1^2} \sum_{m=1}^{l-1} \theta_m(\vec{k}_1) \delta_{l-m}(\vec{k}_2), \tag{16a}$$

$$B_l(\vec{k}) \equiv -\int d^3k_1 \int d^3k_2 \, \delta_D(\vec{k}_1 + \vec{k}_2 - \vec{k}) \frac{k^2(\vec{k}_1 \cdot \vec{k}_2)}{k_1^2 k_2^2} \sum_{m=1}^{l-1} \theta_m(\vec{k}_1) \theta_{l-m}(\vec{k}_2). \tag{16b}$$

Solving equations (15) for $\delta_l$ and $\theta_l$ gives, for $l > 1$,

$$\delta_l(\vec{k}) = \frac{(1+2l)A_l(\vec{k}) - B_l(\vec{k})}{(2l+3)(l-1)}, \quad \theta_l(\vec{k}) = \frac{-3A_l(\vec{k}) + lB_l(\vec{k})}{(2l+3)(l-1)}. \tag{17}$$

Equations (16) and (17) give recursion relations for $\delta_l(\vec{k})$ and $\theta_l(\vec{k})$, with starting values $\delta_1(\vec{k})$ and $\theta_1 = -\delta_1$. The general solution may be written

$$\delta_l(\vec{k}) = \int d^3q_1 \cdots \int d^3q_l \, \delta_D(\vec{q}_1 + \cdots + \vec{q}_l - \vec{k}) F_l(\vec{q}_1, \ldots, \vec{q}_l) \delta_1(\vec{q}_1) \cdots \delta_1(\vec{q}_l), \tag{18a}$$

$$\theta_l(\vec{k}) = -\int d^3q_1 \cdots \int d^3q_l \, \delta_D(\vec{q}_1 + \cdots + \vec{q}_l - \vec{k}) G_l(\vec{q}_1, \ldots, \vec{q}_l) \delta_1(\vec{q}_1) \cdots \delta_1(\vec{q}_l). \tag{18b}$$

From equations (16)–(18) we obtain recursion relations for $F_l$ and $G_l$:

$$F_l(\vec{q}_1, \ldots, \vec{q}_l) = \sum_{m=1}^{l-1} \frac{G_m(\vec{q}_1, \ldots, \vec{q}_m)}{(2l+3)(l-1)} \left[ (1+2l) \frac{\vec{k} \cdot \vec{k}_1}{k_1^2} F_{l-m}(\vec{q}_{m+1}, \ldots, \vec{q}_l) \right.$$
$$\left. + \frac{k^2(\vec{k}_1 \cdot \vec{k}_2)}{k_1^2 k_2^2} G_{l-m}(\vec{q}_{m+1}, \ldots, \vec{q}_l) \right], \tag{19a}$$

$$G_l(\vec{q}_1, \ldots, \vec{q}_l) = \sum_{m=1}^{l-1} \frac{G_m(\vec{q}_1, \ldots, \vec{q}_m)}{(2l+3)(l-1)} \left[ 3 \frac{\vec{k} \cdot \vec{k}_1}{k_1^2} F_{l-m}(\vec{q}_{m+1}, \ldots, \vec{q}_l) \right.$$
$$\left. + l \frac{k^2(\vec{k}_1 \cdot \vec{k}_2)}{k_1^2 k_2^2} G_{l-m}(\vec{q}_{m+1}, \ldots, \vec{q}_l) \right], \tag{19b}$$

where $\vec{k}_1 \equiv \vec{q}_1 + \cdots + \vec{q}_m$, $\vec{k}_2 \equiv \vec{q}_{m+1} + \cdots + \vec{q}_l$, $\vec{k} \equiv \vec{k}_1 + \vec{k}_2$ and $F_1 = G_1 = 1$.

To calculate the power spectrum we shall prefer to use symmetrized forms of $F_l$ and $G_l$, denoted $F_l^{(s)}$ and $G_l^{(s)}$ and obtained by summing the $l!$ permutations of $F_l$ and $G_l$ over their $l$ arguments and dividing by $l!$. Since the arguments are dummy variables of integration the symmetrized functions can be used in equations (18) without changing the result. The recursion relations in equations (19) may be used to compute the power spectrum at any order in perturbation theory. Substituting equation (14) into equation (8), we have

$$\begin{aligned}
P(k,\tau)\delta_D(\vec{k}+\vec{k}') &= \langle \delta(\vec{k},\tau)\,\delta(\vec{k}',\tau)\rangle \\
&= a^2(\tau)\langle \delta_1(\vec{k})\,\delta_1(\vec{k}')\rangle + a^4(\tau)\bigg[\langle \delta_1(\vec{k})\,\delta_3(\vec{k}')\rangle + \langle \delta_2(\vec{k})\,\delta_2(\vec{k}')\rangle \\
&\quad + \langle \delta_3(\vec{k})\,\delta_1(\vec{k}')\rangle\bigg] + O(\delta_1^6) \\
&= \sum_{l=2}^{\infty} a^l(\tau) \sum_{m=1}^{l-1} P_{m,l-m}(k)
\end{aligned} \qquad (20)$$

with $l$ restricted to being an even integer. Equation (20) explicitly shows all the terms contributing to the power spectrum at fourth order in the initial density field $\delta_1$. To distinguish the different terms that contribute to the power spectrum at a given order we have introduced $P_{m,l-m}(k)$, the $m$th contributing term at order $l$, defined as:

$$\langle \delta_m(\vec{k})\,\delta_{l-m}(\vec{k}')\rangle \equiv P_{m,l-m}(k)\,\delta_D(\vec{k}+\vec{k}') \,. \qquad (21)$$

Substituting for $\delta_l$ and $\delta_m$ in equation (21) gives:

$$\begin{aligned}
P_{m,l-m}(k,\tau) = a^l \int d^3q_1\ldots d^3q_{(l-2)}\,\langle \hat{\delta}_1(\vec{q}_1)\ldots \delta_1(\vec{k}-\vec{q}_1-\ldots-\vec{q}_{m-1})\,\delta_1(\vec{q}_m)\ldots \\
\delta_1(\vec{k}-\vec{q}_m-\ldots-\vec{q}_{(l-2)})\rangle\,M_1(\vec{q}_1,\ldots,\vec{q}_{(l-2)},\vec{k})\,,
\end{aligned} \qquad (22)$$

where $M_1$ is a dimensionless function of $F_m^{(s)}$ and $F_{l-m}^{(s)}$.

We can now proceed to demonstrate that in the absence of long wave divergences (i.e., if the contributions to $P(k)$ from low $q$ are convergent), a perturbative expansion can be consistently defined such that the self-similar scaling of equation (2) is obeyed. At sufficiently small scales $\delta\rho/\rho > 1$ even at the earliest times — hence the perturbative expansion breaks down at these scales. This means that even in the absence of a divergence as $k \to \infty$, a high-$k$ cutoff, $k_u$, must be used to truncate the perturbative integrals. The requirement of a cutoff restricts the nonlinear effects that can be studied perturbatively. Nonlinear contributions *from* wavenumbers $q > k_u$ cannot be evaluated, but the contribution *to* any $k$ (from all $q < k_u$) are calculable. We define the high-$k$ cutoff to be time dependent such that: $k_u(\tau) \propto k_{ss}(\tau) \propto a^{-2/(3+n)}$. Once $k_u(\tau)$ is chosen to scale with $k_{ss}(\tau)$, there is no other scaling in the problem, hence it should come as no surprise that the power spectrum takes the self-similar form of equation (3).

Without loss of generality, consider the contribution to $P(k,\tau)$ from the term $P_{m,l-m}(k,\tau)$ defined in equation (21). On taking the ensemble average, the $(l-2)$ independent phase factors contained in the functions $\delta_1(q_i)$ must cancel pairwise for the result to be non-zero (recall that the $\delta_1$'s are taken to be independent Gaussian random variables). Thus we obtain $(l-2)/2$ Dirac-delta functions which reduce the number of integration variables to $(l-2)/2$. There are also $l/2$ powers of $P_{11}(q_i) = Aq_i^n$ present. Collecting the relevant factors which provide the $k$ and $\tau$ dependence, and imposing the high-$k$ cutoff $k_u(\tau)$, we obtain

$$P_{m,l-m}(k,\tau) = a^l \, k^{3(l-2)/2+nl/2} \, A^{l/2} \, M_2(k_u(\tau)/k) \,, \tag{23}$$

where $M_2$ is another dimensionless function. Taking $k_u(\tau) = k_0 a^{-2/(3+n)}$, where $k_0$ is a constant, and introducing a new dimensionless function $M_3$, we finally obtain

$$P_{m,l-m}(k,\tau) = a^{6/(3+n)} \, k_0^{-3} \, M_3(k \, a^{2/(3+n)}/k_0) \,. \tag{24}$$

Equation (24) gives the desired self-similar form for the power spectrum (equation 3), with characteristic scales $k_c(\tau) \propto a^{-2/(3+n)}$ in agreement with equation (2).

Thus at every order in perturbation theory the self-similar scaling of the power spectrum, and therefore of physical measures of perturbation growth such as the smoothed density contrast in real space constructed from it, is preserved. However, this scaling is broken if the perturbative integrals diverge as $k \to 0$, thus requiring a low-$k$ cutoff. This possibility is considered next.

### 3.2. Long Wave Divergences in Perturbative Contributions

The perturbative formalism presented above was used in an earlier work (Jain & Bertschinger 1994) to study the second order power spectrum. It was demonstrated that at second order in the power spectrum there are terms that are divergent for $n < -1$ due to the contribution from $k \to 0$. However, as first shown by Vishniac (1983) the two contributing terms, $P_{22}$ and $P_{13}$, exactly cancel each other at leading order as $k \to 0$ thus keeping the net contribution finite for $n > -3$. This cancellation does not prove there is no divergence in the power spectrum. We must investigate higher-order terms $\delta_l(\vec{k})$. It is tedious to evaluate the full expressions for $\delta_l$ for $l > 2$ and then to form the power spectrum contributions $P_{m,l-m}(k)$. However, we do not need the exact nonlinear power spectrum if we are interested only in determining whether leading-order long wave divergences are canceled. In this case, it is sufficient to work from the outset with only the leading-order divergent parts of $\delta_l(\vec{k})$.

Iterating equations (16) and (17), one finds that the leading-order divergences arise from the term with $m = 1$ in equations (16), with the contribution doubled in equation (16b) because of the term $m = l - 1$. The leading-order divergent contributions are then

$$A_l(\vec{k}) \sim \delta_{l-1}(\vec{k})\,\zeta(\vec{k})\,, \quad B_l(\vec{k}) \sim \theta_{l-1}(\vec{k})\,\zeta(\vec{k})\,, \quad \zeta(\vec{k}) \equiv \int d^3q\,\frac{\vec{k}\cdot\vec{q}}{q^2}\delta_1(\vec{q})\,. \qquad (25)$$

The leading-order divergence appears at $q = 0$ in the function $\zeta(\vec{k})$. Using equation (17) and iterating gives the leading-order divergences of $\delta_l$ and $\theta_l$:

$$\delta_l(\vec{k}) \sim \theta_l(\vec{k}) \sim \frac{\zeta^{l-1}(\vec{k})}{(l-1)!}\,\delta_1(\vec{k})\,. \qquad (26)$$

From equations (21) and (26) we arrive at the leading-order divergent part of $P_{m,l-m}$:

$$P_{m,l-m}(\vec{k}) \sim \frac{(-1)^{m-1}\left\langle \zeta^{l-2}\right\rangle}{(m-1)!\,(l-m-1)!}\,P_{11}(k)\,, \qquad (27)$$

where

$$\left\langle \zeta^{l-2}\right\rangle = (l-3)!!\,k^{l-2}\left[-\int\frac{d^3q}{3q^2}P_{11}(q)\right]^{(l-2)/2}\,. \qquad (28)$$

The net contribution to the leading-order divergent part of the nonlinear power spectrum (20) is

$$P(k,\tau) \sim a^2(\tau)\,P_{11}(k)\sum_{m=2}^{\infty}\frac{a^{l-2}(\tau)\left\langle \zeta^{l-2}\right\rangle}{(l-2)!}\sum_{m=1}^{l-1}\frac{(-1)^{m-1}(l-2)!}{(m-1)!\,(l-m-1)!}\,. \qquad (29)$$

Now, the sum over $m$ is just the binomial expansion of $(1-1)^{l-2}$. Therefore, the sum vanishes for $n > 2$ and the leading-order divergences cancel at every nonlinear order of perturbation theory!

This surprising result does not prove that $P(k,\tau)$ is finite, however. Equation (26) gives only the most divergent contribution to $\delta_l(\vec{k})$, $\zeta^{l-1}$. Terms diverging as $\zeta^{l-2}$ or more slowly have been neglected. The nonlinear power spectrum may still have an $l$-point contribution that diverges as $\langle\zeta^{l-4}\rangle$. The lowest order at which such sub-dominant divergences would appear is $l = 6$: $P = a^6(P_{15} + P_{24} + P_{33} + P_{42} + P_{51})$. Using equations (18a) and (21) gives the integral expressions for the contributing terms:

$$P_{15}(k) = 15 P_{11}(k)\int d^3q_1\,P_{11}(q_1)\int d^3q_2\,P_{11}(q_2)\,F_5^{(s)}(\vec{q}_1,-\vec{q}_1,\vec{q}_2,-\vec{q}_2,\vec{k})\,, \qquad (30)$$

$$P_{24}(k) = 12\int d^3q_1\ P_{11}(q_1)\int d^3q_2 P_{11}(q_2)\,P_{11}(|\vec{k}-\vec{q}_2|)\,F_4^{(s)}(\vec{q}_1,-\vec{q}_1,\vec{q}_2,\vec{k}-\vec{q}_2)$$
$$\times F_2^{(s)}(-\vec{q}_2,\vec{q}_2-\vec{k})\,, \qquad (31)$$

$$P_{33}(k) = 9P_{11}(k)\int d^3q_1\, P_{11}(q_1)\int d^3q_2\, P_{11}(q_2)\, F_3^{(s)}(\vec{q}_1,-\vec{q}_1,\vec{k})\, F_3^{(s)}(\vec{q}_2,-\vec{q}_2,-\vec{k})$$
$$+6\int d^3q_1 P_{11}(q_1)\int d^3q_2\, P_{11}(q_2)\, P_{11}(|\vec{k}-\vec{q}_1-\vec{q}_2|)$$
$$\times F_3^{(s)}(\vec{q}_1,\vec{q}_2,\vec{k}-\vec{q}_1-\vec{q}_2)\, F_3^{(s)}(-\vec{q}_1,-\vec{q}_2,\vec{q}_1+\vec{q}_2-\vec{k})\,. \qquad (32)$$

The factors of 15, 12, 9 and 6 in equations (30)–(32) come from the number of equivalent graphs obtained by relabeling the internal wavevectors, assuming that $F^{(s)}$ and $G^{(s)}$ are fully symmetric in all their arguments. By examining the form of the sub-dominant divergent parts of the contributing terms, it appears that the second term in $P_{33}$ must cancel with the divergent part of $P_{24}$ as $q_1 \to 0$, and the other terms shown must cancel separately if there is to be a net cancellation. With increasing $l$, the full expression for $F_l(\vec{q}_1,\ldots,\vec{q}_l)$ rapidly becomes unwieldy. We did not complete this calculation owing to the computational complexity involved, and the fact that a null result would leave open the possibility of sub-dominant divergences at still higher orders.

The results from the analysis of perturbation theory are therefore not conclusive. The cancellation of leading divergences is certainly suggestive of an underlying kinematical effect which appears in the power counting assessment of the divergence, but cancels out on computing the net dynamical influence on the power spectrum. We will interpret this cancellation by examining the phase of $\delta(\vec{k})$ in Section 4. However it is not feasible to evaluate all the divergent terms at arbitrary order in perturbation theory, therefore we pursue a somewhat different approximation to evaluate long wave mode coupling in the next section.

## 4. Analytic Approximation for Long Wave Mode Coupling

The approach in this section relies on assuming that the nonlinear terms in the Fourier space cosmological fluid equations (8) are dominated by the coupling of long wave modes. With this ansatz the mode coupling contribution is estimated and then checked for self-consistency. This allows us to obtain a leading order solution for the phase shift as described in Section 4.1. To make further progress we need to make the additional assumption that at low $k$, the Fourier space density and velocity fields are continuous and therefore amenable to a Taylor series expansion. This analysis is presented in Section 4.2, and its limitations are discussed.

### 4.1. Solution for the Phase Shift

In equations (8) the integrands on the right-hand side involve products of $\delta$ and $\theta$ evaluated at $\vec{k}_1$ and $(\vec{k} - \vec{k}_1)$. Let

$$\delta(\vec{k} - \vec{k}_1) = \delta(\vec{k}) + \epsilon(\vec{k}, \vec{k}_1) \; ; \; \theta(\vec{k} - \vec{k}_1) = \theta(\vec{k}) + \omega(\vec{k}, \vec{k}_1), \tag{33}$$

where $\epsilon$ and $\omega$ are unknown functions. In this section we shall use "function" to refer to random valued fields as well. We shall also suppress the time dependence of $\delta(\vec{k}, \tau)$ and of $\theta(\vec{k}, \tau)$ for convenience, though when we introduce the linear solutions the $\tau$ dependent part will be explicitly written. Substituting equation (33) into equations (8) gives,

$$\frac{\partial \delta(\vec{k})}{\partial \tau} + \theta(\vec{k}) = - \int d^3 k_1 \, \theta(\vec{k}_1) \frac{\vec{k}_1 \cdot \vec{k}}{k_1^2} \left[ \delta(\vec{k}) + \epsilon(\vec{k}, \vec{k}_1) \right] \equiv A(\vec{k}), \tag{34a}$$

$$\frac{\partial \theta(\vec{k})}{\partial \tau} + \frac{\dot{a}}{a} \theta(\vec{k}) + \frac{6}{\tau^2} \delta(\vec{k}) = - \int d^3 k_1 \, k^2 \frac{\vec{k}_1 \cdot (\vec{k} - \vec{k}_1)}{2 k_1^2 |\vec{k} - \vec{k}_1|^2} \theta(\vec{k}_1) \left[ \theta(\vec{k}) + \omega(\vec{k}, \vec{k}_1) \right] \equiv B(\vec{k}). \tag{34b}$$

In order to estimate the nonlinear effects of long wave modes we assume that the integrands on the right-hand side of equations (34) are dominated by the contribution from $k_1 \ll k$. We then approximate $\theta(\vec{k}_1)$ by the linear solutions given in equation (10): $\theta_1(\vec{k}_1, \tau) = -\dot{a}\delta_1(\vec{k}_1)$, because for $k_1 \ll k$ the amplitude of the density perturbations is taken to be very small. Thus we write the right-hand side of equations (34) as

$$A(\vec{k}) = \dot{a}\, \delta(\vec{k}) \int d^3 k_1 \frac{\vec{k}_1 \cdot \vec{k}}{k_1^2} \delta_1(\vec{k}_1) \; + \dot{a} \int d^3 k_1 \frac{\vec{k}_1 \cdot \vec{k}}{k_1^2} \delta_1(\vec{k}_1) \epsilon(\vec{k}, \vec{k}_1), \tag{35a}$$

$$B(\vec{k}) = \dot{a}\, \theta(\vec{k}) \int d^3 k_1 \frac{\vec{k}_1 \cdot \vec{k}}{k_1^2} \delta_1(\vec{k}_1) \; + \dot{a} \int d^3 k_1 \frac{\vec{k}_1 \cdot \vec{k}}{k_1^2} \delta_1(\vec{k}_1) \omega(\vec{k}, \vec{k}_1). \tag{35b}$$

In the expression for $B(\vec{k})$ we have multiplied the right-hand side by 2 to include the contribution from $(\vec{k} - \vec{k}_1) \to 0$ as required by the symmetry of the integrand. We have also explicitly written out the $\tau$ dependence of $\theta(\vec{k}_1)$, so that $\delta_1(\vec{k}_1)$ does not depend on $\tau$. We now define the integrals:

$$\vec{\alpha} = -i \int d^3 k_1 \frac{\vec{k}_1}{k_1^2} \delta_1(\vec{k}_1), \tag{36}$$

where $i = \sqrt{-1}$; and,

$$E(\vec{k}) = \dot{a} \int d^3 k_1 \frac{\vec{k} \cdot \vec{k}_1}{k_1^2} \delta_1(\vec{k}_1) \epsilon(\vec{k}, \vec{k}_1) \; ; \; W(\vec{k}) = \dot{a} \int d^3 k_1 \frac{\vec{k} \cdot \vec{k}_1}{k_1^2} \delta_1(\vec{k}_1) \omega(\vec{k}, \vec{k}_1). \tag{37}$$

Using these definitions equations (34) can be written as:

$$\frac{\partial \delta(\vec{k})}{\partial \tau} + \theta(\vec{k}) = i\, \dot{a}\, \vec{k} \cdot \vec{\alpha}\, \delta(\vec{k}) + E(\vec{k}), \tag{38}$$

$$\frac{\partial \theta(\vec{k})}{\partial \tau} + \frac{\dot{a}}{a} \theta(\vec{k}) + \frac{6}{\tau^2} \delta(\vec{k}) = i \dot{a} \vec{k} \cdot \vec{\alpha} \, \theta(\vec{k}) + W(\vec{k}) \,. \tag{39}$$

The above equations are exact aside from using the linear solutions for $\theta(\vec{k}_1)$ in the right-hand side of equations (34). We have defined $\vec{\alpha}$ in equation (36) so that it is purely real. This can be verified by using the relation of $\delta_1$ to its complex conjugate: $\delta_1(\vec{k}_1) = \delta_1^*(-\vec{k}_1)$, which is required to ensure that $\delta(\vec{x})$ is real. The reason for introducing $\vec{\alpha}$ is that it is independent of $\vec{k}$ and $\tau$, therefore, given the initial density $\delta_1$, it can be treated as a numerical constant.

We now turn to the issue of long wave divergences. The variance of $\vec{\alpha}$ is:

$$\langle \alpha^2 \rangle \equiv \langle \vec{\alpha} \cdot \vec{\alpha} \rangle = \int d^3k_1 \frac{P_{11}(k_1)}{k_1^2} = 4\pi \int dk_1 P_{11}(k_1) \,. \tag{40}$$

Thus $\langle \alpha^2 \rangle$ is a divergent integral for $n < -1$. To proceed further we need to estimate the degree of divergence of the integrals $E(\vec{k})$ and $W(\vec{k})$. We do so by using equation (33) to substitute for $\epsilon$ and $\omega$ in $E$ and $W$. The resulting expression for the variance of $E$ is:

$$\langle |E(\vec{k})|^2 \rangle = \dot{a}^2 \int d^3k_1 \int d^3k_2 \left(\frac{\vec{k} \cdot \vec{k}_1}{k_1^2}\right) \left(\frac{\vec{k} \cdot \vec{k}_2}{k_2^2}\right)$$
$$\times \left\langle \delta_1(\vec{k}_1) \delta_1^*(\vec{k}_2) \left[\delta(\vec{k}-\vec{k}_1) - \delta(\vec{k})\right] \left[\delta^*(\vec{k}-\vec{k}_2) - \delta^*(\vec{k})\right] \right\rangle \,. \tag{41}$$

To simplify this expression we assume that for the purpose of assessing the degree of divergence, all the fields involved are well approximated by the linear solution. Then the expectation values can be evaluated using the properties of Gaussian random fields. Of the twelve terms that result, the leading contribution in the long wave limit arises from the term with $\langle \delta_1(\vec{k}_1) \delta_1^*(\vec{k}_2) \rangle \langle \delta(\vec{k}-\vec{k}_1) \delta^*(\vec{k}-\vec{k}_2) \rangle$ in the integrand. This contribution is:

$$\langle |E(\vec{k})|^2 \rangle \sim \frac{1}{3} \dot{a}^2 \, P(k) \, \delta_D(0) \, k^2 \alpha^2 \,. \tag{42}$$

The variance of the first term on the right-hand side of equation (38), $i \dot{a} \vec{\alpha} \cdot \vec{k} \delta(\vec{k})$ is exactly the same as the above result for $\langle |E(\vec{k})|^2 \rangle$, hence both terms must be retained at the same order in evaluating the long wave contribution. Likewise, it is easy to show that $W$ in equation (39) is of the same order as the first term on the right-hand side, and is also proportional to $\alpha$ in its degree of divergence.

Equations (38) and (39) can be re-written as a pair of second order differential equations in $\tau$ for $\delta$ and $\theta$. For $\delta$ the result is, with $\tilde{\alpha} \equiv \vec{k} \cdot \vec{\alpha}$,

$$\ddot{\delta} + \dot{\delta}\left(-2i\dot{a}\,\tilde{\alpha} + \frac{\dot{a}}{a}\right) + \delta\left(-\dot{a}^2\,\tilde{\alpha}^2 - 3i\ddot{a}\tilde{\alpha} - \frac{6}{\tau^2}\right) - \frac{\dot{a}}{a}E + i\dot{a}\tilde{\alpha}E + W - \dot{E} = 0\,. \tag{43}$$

Since, to leading order, we know the degree of divergence of the variances of all the terms involved in this equation, we are now in a position to evaluate the effect of long wave divergences. The variables in equation (43) are complex, hence it can be simplified further by separating the real and imaginary parts. To this end we express $\delta$ in terms of its amplitude and phase as:

$$\delta(\vec{k}, \tau) = \Delta(\vec{k}, \tau) \, e^{i\phi(\vec{k}, \tau)} \, . \tag{44}$$

For convenience, we further define $\bar{E} \equiv E e^{-i\phi}$, $\bar{W} \equiv W e^{-i\phi}$, and $\bar{\dot{E}} \equiv \dot{E} e^{-i\phi}$. With these substitutions equation (43) separates into its real and imaginary parts (respectively) as:

$$\ddot{\Delta} + \frac{\dot{a}}{a}\dot{\Delta} + \Delta\left(-\dot{\phi}^2 + 2\dot{\phi}\dot{a}\tilde{\alpha} - \dot{a}^2\tilde{\alpha}^2 - \frac{6}{\tau^2}\right) + \text{Re}\left[\frac{\dot{a}}{a}\bar{E} + i\dot{a}\tilde{\alpha}\bar{E} + \bar{W} - \bar{\dot{E}}\right] = 0, \tag{45a}$$

$$\dot{\Delta}\left(2\dot{\phi} - 2\dot{a}\tilde{\alpha}\right) + \Delta\left(\ddot{\phi} + \frac{\dot{a}}{a}\dot{\phi} - 3\ddot{a}\tilde{\alpha}\right) + \text{Im}\left[\frac{\dot{a}}{a}\bar{E} + i\dot{a}\tilde{\alpha}\bar{E} + \bar{W} - \bar{\dot{E}}\right] = 0. \tag{45b}$$

"Re" and "Im" denote the real and imaginary parts, respectively, of the expressions in the square brackets.

We now make the ansatz that $\phi \sim O(\alpha)$, and that $\Delta \sim O(\alpha^p)$, where $0 \leq p < 1$. Since $E \sim W \sim O(\alpha)$ (in an rms sense) from equation (42), keeping terms of $O(\alpha^2)$ in equation (45a) gives,

$$\Delta\left(-\dot{\phi}^2 + 2\dot{\phi}\dot{a}\tilde{\alpha} - \dot{a}^2\tilde{\alpha}^2\right) + \text{Re}\left[i\dot{a}\tilde{\alpha}\bar{E} - \bar{\dot{E}}\right] = 0. \tag{46}$$

As we shall see below, retaining the term $\bar{\dot{E}}$ is required for consistency. We make the assumption that at leading order in $\alpha$ the two parts of equation (46) in brackets vanish separately (this will also be justified below). The first part gives a quadratic equation for $\dot{\phi}$,

$$\dot{\phi}^2 - 2\dot{\phi}\dot{a}\tilde{\alpha} + \dot{a}^2\tilde{\alpha}^2 = 0, \tag{47}$$

which has the solution, $\dot{\phi} = \dot{a}\tilde{\alpha} = \dot{a}\vec{k} \cdot \vec{\alpha}$. Thus the leading order solution for $\phi$ is:

$$\phi(\vec{k}, \tau) = a(\tau)\, \vec{k} \cdot \vec{\alpha} + \phi_i(\vec{k}), \tag{48}$$

where $\phi_i(\vec{k})$ is the value of the phase at the initial time.

The solution of equation (48) can be used to justify the assumptions that have been made. Firstly, $\phi$ is indeed of $O(\alpha)$, as assumed at the outset. Further, equations (37) and (48) can be used to simplify the expression for $\bar{E}$ and thereby justify setting the first part of equation (46) to 0 separately. To start with let us write $E$ in terms of $\Delta$ and $\phi$:

$$E = \dot{a} \int d^3k_1 \frac{\vec{k} \cdot \vec{k}_1}{k_1^2} \Delta_1(\vec{k}_1) e^{i\phi(\vec{k}_1)} \left[\Delta(\vec{k} - \vec{k}_1) e^{i\phi(\vec{k} - \vec{k}_1)} - \Delta(\vec{k}) e^{i\phi(\vec{k})}\right]. \tag{49}$$

Differentiating equation (49) with respect to $\tau$ and multiplying by $e^{-i\phi(\vec{k})}$ gives,

$$\dot{\bar{E}} = \frac{\ddot{a}}{\dot{a}}\bar{E} + \dot{a}\int d^3k_1 \frac{\vec{k}\cdot\vec{k}_1}{k_1^2}\Delta_1(\vec{k}_1)e^{i\phi(\vec{k}_1)}\left[\left\{\dot{\Delta}(\vec{k}-\vec{k}_1) + i\Delta(\vec{k}-\vec{k}_1)\dot{\phi}(\vec{k}-\vec{k}_1)\right\}\right.$$
$$\left.\times e^{i\phi(\vec{k}-\vec{k}_1)-i\phi(\vec{k})} - \left\{\dot{\Delta}(\vec{k}) + i\Delta(\vec{k})\dot{\phi}(\vec{k})\right\}\right]. \tag{50}$$

The leading order terms on the right-hand side above are the two terms with $\dot{\phi}$: they are at least of $O(\alpha^2)$. However, by substituting $\dot{\phi} = \dot{a}\tilde{\alpha}$ into equation (50), and comparing with the expression for $\bar{E}$ that follows from equation (49), it can be seen that these leading order terms exactly cancel the contribution from $i\dot{a}\tilde{\alpha}\bar{E}$ in equation (46). Thus the surviving terms in the second part of equation (46) are all of lower order than $O(\alpha^2)$ — therefore they can be neglected in comparison to the first part of the equation which was used to get the solution for $\phi$ of equation (48). This establishes the consistency of the approximations used to obtain this solution.

The variance of the phase shift given by equation (48) is:

$$\left\langle\left[\delta\phi(\vec{k},\tau)\right]^2\right\rangle = a(\tau)^2\, k^2 \int d^3k_1\, \frac{(\hat{k}\cdot\hat{k}_1)^2}{k_1^2}\, P_{11}(k_1), \tag{51}$$

where $\delta\phi(\vec{k},\tau) \equiv \phi(\vec{k},\tau) - \phi_i(\vec{k})$, and $\hat{k}$ and $\hat{k}_1$ are unit vectors. Since $\vec{k}$ is a fixed external vector the angular integral can be performed so that the final result depends only on the magnitude of $\vec{k}$:

$$\left\langle\left[\delta\phi(\vec{k},\tau)\right]^2\right\rangle = \frac{4\pi}{3}a(\tau)^2\, k^2 \int dk_1\, P_{11}(k_1). \tag{52}$$

Thus the leading order solution for $\delta(\vec{k})$ involves a growing (and, for $n < -1$, divergent) phase shift, but there are no contributions to the amplitude at this order. The above analysis can be repeated for the velocity divergence $\theta(\vec{k})$ to verify that the leading order result for $\theta(\vec{k})$ is the same, with equation (48) giving the solution for its phase as well. These results were obtained by retaining terms of $O(\alpha^2)$. Since divergent terms of $O(\alpha)$ are also present in the equations we cannot say anything conclusive about the degree of (possible) divergence of the amplitude $\Delta(\vec{k})$. In the following section we shall address this question by expanding the equations to next order in $\alpha$ with some additional assumptions.

The solution (48) for the phase shift has a simple physical interpretation. As noted in Section 2, the linear bulk velocity $v_b$ diverges due to contributions from long wave modes (equation 13). The limiting form of the integral given in equation (13) for $v_b^2$, and that of equation (52), is the same. The connection between them can be made more precise by imagining a single sine-wave density perturbation in real space: $\delta(\vec{x},\tau) = \delta_o\sin(\vec{k}\cdot\vec{x})$. Now suppose that the fluid in which this perturbation is made is moved with a uniform translational velocity of magnitude $v_b(\tau)$ given by (13) (the scale $x$ in equation 13 has no connection to the spatial variable $\vec{x}$ used here). The distance moved by each fluid element is $\int d\tau\, v_b(\tau) = v_b(\tau)\, a(\tau)/\dot{a}(\tau)$, where we have used $v_b(\tau) \propto \dot{a}(\tau)$. If the coordinate frame is kept fixed relative to this translational motion, then the density perturbation will acquire the following time dependence due to the bulk velocity: $\delta(\vec{x},\tau) = \delta_o \sin[\vec{k}\cdot(\vec{x} - \hat{e}\, v_b a/\dot{a})]$, where $\hat{e}$ is the direction of the bulk velocity. Therefore $\delta$ acquires a phase shift: $\delta\phi(\vec{k}) = \vec{k}\cdot\hat{e}\, a/\dot{a}\, v_b$. On squaring, and averaging over angles between $\vec{k}$ and $\hat{e}$, this gives:

$$\left\langle \left[\delta\phi(\vec{k},\tau)\right]^2 \right\rangle = \frac{1}{3} \frac{a^2}{\dot{a}^2} k^2\, v_b^2(\tau). \tag{53}$$

Note that the averaging over angles is consistent with the angular integral done to get equation (52), and amounts to estimating the typical phase shift due to a superposition of bulk flows of magnitude $v_b$ directed randomly with respect to $\vec{k}$. Substituting for $v_b^2$ from equation (13) and assuming it is dominated by the contribution from low $k$, we recover the result in equation (52). Thus we have shown that for $n < -1$ the dominant phase shift is due to the kinematical effect of the bulk motion on small scales imparted by long wave modes. This is consistent with the connection between divergences in the nonlinear terms in the fluid equations and $v_b$ made in Section 2.

### 4.2. Taylor Series Expansion

In this section we make an additional assumption about the $\vec{k}$-dependence of $\delta(\vec{k})$ and $\theta(\vec{k})$: we assume that in a small neighborhood around $\vec{k}$, $\delta$ and $\theta$ are smooth, differentiable functions of $\vec{k}$. With this assumption we expand the nonlinear integrals in equations (8) in a Taylor series in $(k_1/k)$ about 0 and restrict the range of integration to small $k_1$. Thus we write: $\delta(\vec{k} - \vec{k}_1) \simeq \delta(\vec{k}) - \vec{k}_1 \cdot \partial\delta/\partial\vec{k}$, and likewise for $\theta(\vec{k} - \vec{k}_1)$. Unfortunately, the standard assumption about $\delta$ and $\theta$ in cosmology is that they are Gaussian random fields at the initial time. Thus at each value of $\vec{k}$ they are given by a random number drawn from a probability distribution. The distribution of $\delta$ or $\theta$ with respect to $\vec{k}$ is quite the opposite of a smooth function, because its values at any two $\vec{k}$ are uncorrelated. We return to this point later in this section, but here we proceed with the Taylor series approach.

With the Taylor expansion described above, the right-hand sides of equations (8), denoted by $C(\vec{k})$ and $D(\vec{k})$, take the form:

$$C(\vec{k}) = -\int d^3k_1 \left[\theta(\vec{k}_1)\frac{\vec{k}_1\cdot\vec{k}}{k_1^2}\left(\delta(\vec{k}) - \vec{k}_1\cdot\frac{\partial\delta(\vec{k})}{\partial\vec{k}}\right) + \theta(\vec{k})\delta(\vec{k}_1) + \ldots\right], \tag{54a}$$

$$D(\vec{k}) = -\int d^3k_1\,\theta(\vec{k}_1) \left[\frac{\vec{k}_1 \cdot \vec{k}}{k_1^2}\theta(\vec{k}) + \left(\frac{2(\vec{k}_1 \cdot \vec{k})^2}{k_1^2 k^2} - 1\right)\theta(\vec{k}) - \frac{\vec{k}_1 \cdot \vec{k}}{k_1^2}\vec{k}_1 \cdot \frac{\partial \theta(\vec{k})}{\partial \vec{k}} + \ldots \right], \tag{54b}$$

where both equations have been expanded to the same order. In equations (54) we have included the contributions from $(\vec{k} - \vec{k}_1) \to 0$ as well. We now write equations (54) at the order shown as linear equations by approximating $\delta$ and $\theta$ at small $k_1$ by the linear solutions $\delta_1$ and $\theta_1$. Recall that we had obtained one linear term on the right-hand side of each equation in the previous section by introducing the integral $\vec{\alpha}$. Here we introduce three new integrals: $\eta$, $\gamma$ and $g_{ij}$,

$$\eta = \int d^3k_1 \delta(\vec{k}_1)\;;\; \gamma = \int d^3k_1[2(\hat{k} \cdot \hat{k}_1)^2 - 1]\delta(\vec{k}_1)\;;\; g_{ij} = \int d^3k_1 \hat{k}_{1i}\hat{k}_{1j}\delta(\vec{k}_1), \tag{55}$$

where $k_{1i}$ and $k_{1j}$ denote the $i$th and $j$th components of the vector $\vec{k}_1$, so that $g_{ij}$ is a tensor. Note that aside from the dependence of $\gamma$ on the direction of $\vec{k}$, all the integrals in equation (55) are independent of $\vec{k}$ and $\tau$. In addition, all the integrals are convergent in an rms sense for $n > -3$.

We proceed by writing down a second order differential equation for $\delta$ in terms of $C$ and $D$:

$$\ddot{\delta} + \frac{\dot{a}}{a}\dot{\delta} - \frac{6}{\tau^2}\delta = \dot{C} + \frac{\dot{a}}{a}C + D. \tag{56}$$

We then use the definitions of equation (55) to rewrite equations (54) as:

$$C(\vec{k}) = i\dot{a}\vec{k} \cdot \vec{\alpha}\,\delta + \dot{a}[k_i g_{ij}\partial_j]\delta - a\eta\theta, \tag{57}$$

and

$$D(\vec{k}) = -i\dot{a}\vec{k} \cdot \vec{\alpha}\,\theta - \dot{a}[k_i g_{ij}\partial_j]\theta - \dot{a}\gamma\theta, \tag{58}$$

where $\partial_j \equiv \partial/\partial k_j$, and the repeated indices $i$ and $j$ are summed over. We will now attempt to solve these equations for the amplitude and phase of $\delta$ to a given order in $\alpha$. We begin by using equations (57) and (58) to eliminate $\theta$ in the terms on the right-hand side of equation (56) (we will also need to use the left-hand sides of equations (8)). Some algebra yields the following equation for $\delta$:

$$\ddot{\delta} + \frac{\dot{a}}{a}\left[\frac{1}{1+a\eta} - \gamma a - 2ia\vec{k} \cdot \vec{\alpha} - 2a(k_i g_{ij}\partial_j)\right]\dot{\delta} - \frac{6}{\tau^2}(1+a\eta)\delta$$
$$-\dot{a}^2\left[\vec{k} \cdot \vec{\alpha} - i(k_i g_{ij}\partial_j \ln \delta)\right]^2 \delta - i\dot{a}^2\left[-k_i g_{ij}\alpha_j + i(k_i g_{ij}\partial_j)^2 \ln \delta\right]\delta$$
$$-i\frac{\dot{a}^2}{a}\left[\vec{k} \cdot \vec{\alpha} - i(k_i g_{ij}\partial_j \ln \delta)\right]\left[\frac{1}{1+a\eta} - \gamma a + \frac{\ddot{a}a}{\dot{a}^2}\right]\delta = 0. \tag{59}$$

We now make a WKB analysis, which relies on taking the phase to be more rapidly varying than the amplitude, to obtain self-consistent equations for the amplitude and phase. After some algebra we get the following relation for $\dot\phi$ by solving equation (59) at $O(\alpha^2)$:

$$\dot\phi = \dot a(\vec k \cdot \alpha + k_i g_{ij}\partial_j\phi). \tag{60}$$

This yields the solution,

$$\phi = k^T F\alpha + \phi_i \ ; \ F = e^{ag}g^{-1}(I - e^{-ag}), \tag{61}$$

where $\vec\alpha$ has been represented as a column vector, $k^T$ denotes a row vector representing $\vec k$, and $g$ and $F$ are 3 by 3 matrices, with $I$ being the identity matrix. This solution can be verified by substitution using equation (59). Note that for $ag < 1$, $F$ can be expanded as a Taylor series: $F \simeq a + a^2g/2 + a^3g^2/6 + ....$ For $ag \ll 1$ the leading order solution is $\phi = a\vec k \cdot \vec\alpha + \phi_i$, in agreement with (48). The solution for $\phi$ in equation (61) can be used in equation (59) to obtain an equation for $\Delta$ only. After some algebra, this equation simplifies to:

$$\ddot\Delta + \frac{\dot a}{a}\left[\frac{1}{1+a\eta} - \gamma a - -2a(k_i g_{ij}\partial_j)\right]\dot\Delta - \frac{6}{\tau^2}(1+a\eta)\Delta + \dot a^2 \Delta(k_i g_{ij}\partial_j \ln\Delta)^2$$
$$+\dot a^2 \Delta(k_i g_{ij}\partial_j)^2\ln\Delta - \frac{\dot a^2}{a}(k_i g_{ij}\partial_j \ln\Delta)\left[\frac{1}{1+a\eta} - \gamma a + \frac{\ddot a a}{\dot a^2}\right]\Delta = 0. \tag{62}$$

Note that in this equation for $\Delta$, all terms involving $i$ and the divergent integral $\alpha$ have canceled out exactly! Hence the solution for $\Delta$ has no dependence on $\alpha$, the only divergent integral in equation (59). Obtaining the full solution for $\Delta$ is still not possible as it requires solving equation (62), a nonlinear partial differential equation; however, for our purpose the key goal was the assessment of the $\alpha$-dependence of $\Delta$. Thus the Taylor series approach leads to two striking results: the solution for the phase given by equation (61), and the result that the evolution of the amplitude is not influenced by any divergent phase integrals.

The above conclusions thus support the interpretation discussed in Section 2 that for $-3 < n < -1$, there is no dynamically relevant divergence. Hence the evolution of scale free spectra will obey the standard self-similar scaling provided the statistics used are relevant to the growth of density perturbations. The divergent growth of the phase is not a measure of perturbation growth as it arises from bulk motions. However these conclusions rest on the assumption of a valid Taylor series approximation for $\delta$ and $\theta$. This assumption cannot be justified in the cosmological context for random-phase Gaussian initial conditions. It can be argued that the Taylor series expansion becomes a reasonable approximation at sufficiently late times when nonlinear evolution reduces the stochastic character of the density field. However this is at best a qualitative argument, and we must therefore regard the conclusions of this section as being suggestive of the answer, but not proven results.

In Paper II we shall measure the growth of the amplitude and phase of the density field in scale free simulations. The results from an $n = -2$ simulation in particular will allow us to test the analytical results of this section and to assess the validity of our approximations.

## 5. Summary

The goal of this paper was to examine the self-similar scaling of initially scale free cosmological spectra, $P(k) \propto k^n$. We emphasized that the scaling properties for $n < -1$ have not been adequately studied either analytically or through N-body simulations. Indeed some results from N-body simulations suggested that the scaling properties of $n = -2$ simulations were different from those with $n \geq -1$. In this paper we have used analytical techniques to investigate the possible breaking of self-similar scaling for $n < -1$.

We motivated this investigation by demonstrating that terms diverging due to the contribution from long wave modes were present in the cosmological fluid equations, but arguing that a detailed analysis was needed to assess their dynamical influence on the growth of density perturbations. In Section 3 we examined perturbative contributions to the power spectrum to examine the possibility of long wave divergences in these contributions for $-3 < n < -1$. We found that divergent terms were indeed present, but that the leading order divergences canceled out at each order in perturbation theory. Terms which diverged less strongly were also present, but due to the computational complexity involved we did not proceed further with the perturbative analysis. We developed a non-perturbative approximation to study the nonlinear coupling of long wave modes in Section 4. We obtained a solution for the phase shift of the Fourier space density $\delta(\vec{k}, a)$ which is divergent for $-3 < n < -1$. This divergence was interpreted as arising from the kinematical effect of the bulk flows induced by long wave modes. With additional assumptions requiring that $\delta(\vec{k}, a)$ be amenable to a Taylor series expansion around $\vec{k}$, we showed that the evolution of the amplitude of $\delta(\vec{k}, a)$ is not influenced by the divergent terms. It was emphasized that the Taylor series expansion cannot be justified for Gaussian random fields, and therefore the conclusion about the amplitude cannot be regarded as a proof.

Thus the two approaches in this paper strongly suggest that the self-similar scaling of the amplitude of density perturbations holds for $-3 < n < 4$. The subtleties that arise due to long wave divergences for $n < -1$ affect statistical measures such as the rms phase shift, which does not measure dynamical nonlinearity. The general lesson is that certain statistical measures are susceptible to kinematical as well as dynamical influences, and must therefore be interpreted with caution. These include the rms particle displacement and any other statistic that involves the bulk velocity or measures driven by it. In Paper II we study the scaling properties of scale free spectra in N-body simulations and make comparisons with the analytical results of this paper.

It is a pleasure to thank Alan Guth for many stimulating discussions which helped to clarify and sharpen our arguments. We also acknowledge useful discussions with Neal Katz, Samir Mathur, Mark Metzger, Rajaram Nityananda, David Weinberg, Rien van de Weygaert and Simon White.